\documentclass[aps, twocolumn, notitlepage, prl]{revtex4-1}

\makeatletter
\AtBeginDocument{\let\LS@rot\@undefined}
\makeatother

\usepackage{subfigure}
\usepackage{textcomp}
\usepackage{graphicx}
\usepackage{amssymb}
\usepackage{amsmath}
\usepackage{bm}
\usepackage{amsmath, amsthm, amssymb}
\usepackage{dsfont}
\usepackage[colorlinks=true,linkcolor=blue,urlcolor=blue,citecolor=blue]{hyperref}
\usepackage{multirow}
\usepackage{url}
\usepackage{bbold}
\usepackage{blkarray}
\usepackage{array}

\newcolumntype{M}[1]{>{\centering\arraybackslash}m{#1}}
\newcolumntype{N}{@{}m{0pt}@{}}

 \makeatletter
\newsavebox{\@brx}
\newcommand{\llangle}[1][]{\savebox{\@brx}{\(\m@th{#1\langle}\)}%
  \mathopen{\copy\@brx\kern-0.5\wd\@brx\usebox{\@brx}}}
\newcommand{\rrangle}[1][]{\savebox{\@brx}{\(\m@th{#1\rangle}\)}%
  \mathclose{\copy\@brx\kern-0.5\wd\@brx\usebox{\@brx}}}
\makeatother

\def\beq{\begin{equation}}
\def\eeq{\end{equation}}
\def\bea{\begin{eqnarray}}
\def\eea{\end{eqnarray}}

\usepackage[usenames, dvipsnames]{color}

\usepackage{pdfpages} 
\makeatletter
\AtBeginDocument{\let\LS@rot\@undefined}
\makeatother

\begin{document}
\title{Integrability breaking from backscattering}
\author{Javier Lopez-Piqueres}
\affiliation{Department of Physics, University of Massachusetts, Amherst, MA 01003, USA}
\author{Romain Vasseur}
\affiliation{Department of Physics, University of Massachusetts, Amherst, MA 01003, USA}

\begin{abstract}

We analyze the onset of diffusive hydrodynamics in the one-dimensional hard-rod gas subject to stochastic backscattering. While this perturbation breaks integrability and leads to a crossover from ballistic to diffusive transport, it preserves infinitely many conserved quantities corresponding to even moments of the velocity distribution of the gas. In the limit of small noise, we derive the exact expressions for the diffusion and structure factor matrices, and show that they generically have off-diagonal components in the presence of interactions. We find that the particle density structure factor is non-Gaussian and singular near the origin, with a return probability showing logarithmic deviations from diffusion. 

\end{abstract}

\maketitle
\textit{\textbf{Introduction ---}} Hydrodynamics describes the approach from local to global thermal equilibrium in generic many-body systems~\cite{landau2013fluid,spohn2012large}. While it is expected that in the absence of Galilean or Lorentz symmetry, chaotic systems should display diffusive hydrodynamics at long enough times, one dimensional systems can show nontrivial dynamics as a result of proximity to integrability, leading to ballistic, and under some circumstances even superdiffusive or subdiffusive transport~\cite{PhysRevLett.103.216602, PhysRevB.83.035115, PhysRevLett.113.147205,PhysRevB.96.081118, ljubotina2017spin, PhysRevB.97.045407, gopalakrishnan2019kinetic, schemmer2019generalized, bulchandani2020superdiffusive, de2021correlation, bertini2021finite,de2021stability, malvania2021generalized, scheie2021detection, Jepsen:2020aa,de2022subdiffusive, wei2022quantum, bouchoule2022generalized,PhysRevB.106.094303,zechmann2022tunable,peng2022exploiting}. While recent advancements have provided a cohesive theoretical understanding of the hydrodynamics of integrable systems based on stable quasiparticles under the framework of generalized hydrodynamics (GHD) \cite{PhysRevX.6.041065,PhysRevLett.117.207201,bastianello2022introduction}, dynamics away from these fine-tuned points still remain elusive. For small perturbations away from integrability it is believed that generically there will be a crossover from ballistic transport at short enough time scales to conventional ({\em i.e. diffusive}) transport at the longest time scales~\cite{bastianello2021hydrodynamics}. This is a general result that comes from an agnostic approach to the collision integral based on perturbation theory and Fermi's golden rule \cite{friedman2020diffusive, durnin2020non}. While the collision integral and resulting dynamics can be studied analytically in great detail for certain integrability breaking perturbations, such as atom losses~\cite{bouchoule2020effect,PhysRevE.103.042121} and smoothly varying noise~\cite{PhysRevB.102.161110}, most often the best approach to the problem is through a combination of phenomenological insights and sophisticated numerics~\cite{PhysRevB.103.L060302, PhysRevLett.126.090602, PhysRevLett.125.180605,panfil2022thermalization}. The main difficulty can be traced back~\cite{bastianello2021hydrodynamics} to evaluating the matrix elements of the integrability breaking perturbation in generic generalized equilibrium states, also called ``form-factors'', a daunting task that can only be performed for small-momentum transfer perturbations or on finite small-scale systems~\cite{de2018particle, cortes2019thermodynamic, cortes2020generalized, gohmann2017thermal, kitanine2011form}.

\begin{figure}
\includegraphics[scale=0.6, width=\linewidth]{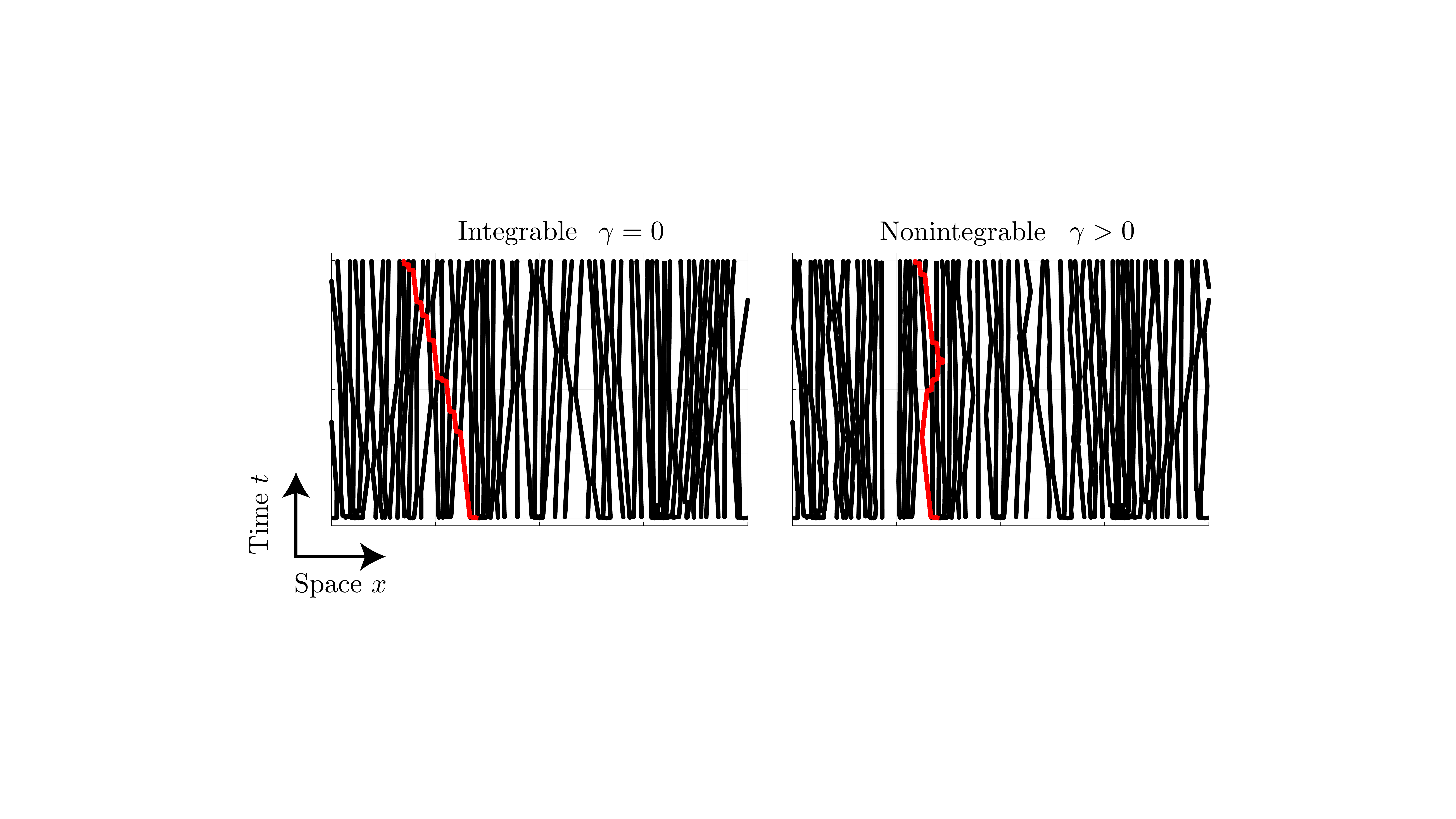}
\caption{\textbf{Snapshots of the dynamics of hard-rods.} Left panel: integrable limit. Right panel: nonintegrable dynamics with backscattering at a rate $\gamma>0$. In red, trajectories of quasiparticles. In the integrable limit, the velocity of quasiparticles gets renormalized as a result of collisions with other quasiparticles. Same initial conditions in both panels.} 
\label{fig_snapshot}
\end{figure}

In this work we address the fate of transport in one of the simplest integrable models in one dimension, the classical hard-rod gas~\cite{percus1969exact, boldrighini1983one, doyon2017dynamics,PhysRevLett.120.164101}, subject to noisy backscattering perturbations \textit{i.e.} stochastic perturbations that reverse the momentum of particles -- and thus correspond to large momentum transfer. While we focus on the classical hard rod gas for concreteness, we note that the hydrodynamics of all known integrable systems, quantum or classical, can be mapped onto {\em generalized} hard-rod gases~\cite{doyon2018soliton}, so our conclusions directly generalize to other models. Stochastic backscattering leads to decay of infinitely many conserved charges, including momentum, but also preserves infinitely many residual conserved quantities corresponding to even moments of the velocity distribution of the gas. The resulting model thus displays features of both integrable and chaotic dynamics. In Fig. \ref{fig_snapshot} we show snapshots of what the dynamics of the hard-rod gas looks like at the integrable point, as well as in the presence of noisy backscattering. The main results of this work are a derivation of the exact expressions for the diffusion and structure-factor matrices of this model. In doing so, we show that the rod density structure factor is highly non-Gaussian and singular as a result of the infinitely-many residual conservation laws.

\textbf{\textit{Hard-rod gas with stochastic backscattering ---}} The one-dimensional hard-rod gas is an integrable model that can be best understood as a set of classical particles subject to a hard-core repulsive potential 
\begin{equation}
\mathcal{H}_0=\sum_{j=1}^N \frac{p_j^2}{2}+\sum_{j<k}U(x_j-x_k), \hspace{0.1in} U(\delta x) =
\begin{cases}
0, & \hspace{0.05in} |\delta x| > a, \\
\infty, & \hspace{0.05in} |\delta x| \leq a,
\end{cases}
\end{equation}
where $a$ denotes the rods' length, and $x_j$ and $p_j$ denote positions and momenta (setting mass $m=1$). Starting from a configuration with $x_{j+1}-x_j \geq  a$, the rods evolve freely until they encounter another rod, $x_{j+1}-x_j=a$, at which point the two rods exchange velocity instantaneously. Because of the simple kinematics of such elastic collisions, the full distribution of velocities (or momenta) is conserved by the evolution and the model is thus integrable. Quasiparticles can be defined by tagging rods with fixed momenta (see Fig.~\ref{fig_snapshot}). Quasiparticles are displaced by an amount $a$ after each collision, so that they move with an effective velocity that depends on the density of all other rods with different momenta. The large-scale, coarse-grained dynamics of hard-rods is described by a Boltzmann-type equation for the phase-space density $\rho_k(x,t)=\frac{d^2N}{dxdk}$ given as~\cite{percus1969exact, boldrighini1983one, doyon2017dynamics} 
\begin{equation} \label{eqGHDrods}
\partial_t \rho +\partial_x(v^{\rm eff}\rho)=0, \hspace{0.05in} v_k^{\rm eff}[\rho]=k+\frac{a\int_{k'}(k-k')\rho_{k'}(x,t)}{1-a\int_{k'}\rho_{k'}(x,t)}.
\end{equation}
This kinetic equation can also be interpreted as an Euler-scale GHD equation for the hard-rod gas~\cite{PhysRevX.6.041065,PhysRevLett.117.207201,doyon2017dynamics}. There are diffusive corrections to this equation, due to the randomness of the scattering shifts arising from thermal fluctuations of the initial state~\cite{lebowitz1968time, spohn1982hydrodynamical, boldrighini1997one, doyon2017dynamics,PhysRevLett.121.160603,gopalakrishnan2018hydrodynamics,de2019diffusion}; in what follows we will ignore those as they are subleading in the limit of weak integrability breaking \cite{friedman2020diffusive}. The integrability of the model can be seen from the infinite set of conservation laws (as $N \to \infty$) corresponding to the various moments w.r.t. the velocities, with charge densities $q_n=\int k^n \rho$. 

We then introduce an integrability-breaking perturbation in the following way: with rate $\gamma$, we stochastically backscatter rods by flipping the sign of their velocity. This perturbation converts right-moving rods into left-moving ones, and {\it vice-versa}. Clearly, this perturbation leads to momentum relaxation, and breaks the conservation of all {\em odd} moments $q_{2n+1}$ of the velocity distribution. On the other hand, all even charges $q_{2n}$ remain conserved: in other words, the odd part of the velocity distribution decays, while the even part remains conserved. Any even velocity distribution is an equilibrium steady-state under this perturbation. 

\textbf{\textit{Generalized Boltzmann equation ---}} In the presence of an integrability breaking perturbation, such as backscattering noise, eq.~(\ref{eqGHDrods}) acquires a right hand side, captured by a collision integral $\mathcal{I}_k[\rho]$. In what follows we shall be interested in the linear response regime, so we write $\rho_k(x,t) \to \rho^*_k + \delta \rho_k(x,t)$, such that the stationary state, $\rho^*$, is an even function of momentum and uniform in space (the latter condition follows from eq. (\ref{eqGHDrods}) subject to $\partial_t \rho^*=0$), $\rho^*_k=n f(k)$, with $n$ the density of particles and $f$ an even function. In this regime the resulting linearized Boltzmann equation  reads~\cite{friedman2020diffusive}
\begin{equation} \label{eq_BE}
\partial_t \delta \rho +A \partial_x \delta \rho = -\Gamma \delta \rho,
\end{equation} 
where $A$ and $\Gamma$ are hydrodynamic matrices that act on velocity space, with $\Gamma_{k,q}\equiv -\delta \mathcal{I}_k/\delta \rho_q|_{\rho=\rho^*}$. The matrix $A$ follows from linearizing~(\ref{eqGHDrods}), and reads~\cite{doyon2017dynamics} $A=R^{-1}v^{\rm eff}R$, with $v_k^{\rm eff} = v_k^{\rm eff}[\rho^*]$, $R=1-\theta^* T$ and $\theta^*=(1-an)^{-1}\rho^*$ an effective occupation number, and the kernel $T$ acts as follows on a test velocity function $(T \psi)_k=-a \int dk' \psi_{k'}$. All matrix operations in those expressions act on velocity space. The operator $\Gamma$ contains the decay rates of the different conserved modes in the original integrable model. Residual conserved quantities thus correspond to zero modes of $\Gamma$. In the case of backscattering noise, we have $(\Gamma \psi)_k=\gamma(\psi_k-\psi_{-k})$. As expected, this perturbation breaks the conservation of odd charges, while preserving the remaining ones. Thus the resulting model is of a new kind, where the system is neither fully chaotic nor integrable: in the following we will show that transport is entirely diffusive, despite the existence of infinitely-many conservation laws. The observable of interest will be the diffusion constant of conserved modes. Since the system under consideration has infinitely-many conserved charges, the resulting diffusion constant will be an infinite dimensional matrix. To derive an expression for this, one can project Eq. (\ref{eq_BE}) onto decaying and conserved modes. The matrix $A$ will mix all modes, so the task is to solve the resulting system of equations. To leading order in a gradient expansion, one can show that the diffusion matrix reads~\cite{suppmat} (see also ~\cite{friedman2020diffusive, durnin2020non})
\begin{equation}
\mathcal{D}=\bar{P}A(P\Gamma P)^{-1}A\bar{P},
\end{equation} 
where $P$ projects onto the subspace of nonconserved modes, and $\bar{P}$ onto its complementary, i.e. onto the subspace of conserved modes. 

\textbf{\textit{Non-interacting limit ---}} To gain some intuition on the problem at hand, we first solve the simple limit of free rods (\textit{i.e.} $a=0$). Intuitively, in that limit each rod is simply undergoing a random walk with mean free path $v_k /(2\gamma)$. 
 In that limit we have $A_{k,k'}=v_k\delta(k-k')$ with $v_k=k$, \textit{i.e.} the velocity of rods in the absence of interactions. The linearized Boltzmann equation simply couples the $(k,-k)$ modes
\begin{equation} \label{eqfreeBE}
\left(\begin{matrix} \partial_t +v_k\partial_x+\gamma & -\gamma \\ -\gamma & \partial_t-v_k\partial_x+\gamma\end{matrix}\right)\left(\begin{matrix} \delta \rho_k \\ \delta \rho_{-k} \end{matrix}\right)=\left(\begin{matrix} 0 \\ 0 \end{matrix} \right).
\end{equation}
Going to Fourier space $(\omega, q)$, this reveals two eigenvalues at low energy: $\omega_q=-i2\gamma+\mathcal{O}(q^2)$ corresponding to the decaying mode $\delta \rho^-_{k}\equiv \delta \rho_k- \delta \rho_{-k}$, and $\omega_q=-i\mathcal{D}q^2+\mathcal{O}(q^4)$, with $\mathcal{D}=v_k^2/(2\gamma)$, corresponding to the diffusive mode $\delta \rho^+_{k} \equiv \delta \rho_k+\delta \rho_{-k}$. Similar equations have been discussed, {\it e.g.}, in the context of the hydrodynamics of stochastic conformal field theories (CFTs)~\cite{bernard2017diffusion}. Away from the free particle limit, the diffusive modes no longer correspond to this particular combination, as the $A$ matrix will connect modes of different velocities $k$. To solve the hard-rod problem with backscattering we take a step back and solve the limit when there are only a discrete number of velocities, in which case $A$ becomes a finite dimensional matrix. 

\textbf{\textit{Discrete velocity distribution ---}} To analyze the case of discrete number of particles it suffices to analyze the case of only two particle species (a more detailed analysis may be found on the Supp. Mat. \cite{suppmat}). Consider a background state given by velocities in the set $\{\pm v_1, \pm v_2\}$, and their respective probabilities $\{\frac{p_1}{2}, \frac{p_2}{2}\}$ with $p_1+p_2=1$. We can write down an exact expression for the discrete version of the hydrodynamic matrices above. These read $T=-aJ_4$, $\Gamma=\gamma \Gamma_1 \oplus \Gamma_2$, with $J_4$ the $4 \times 4$ matrix of all ones, and $\Gamma_i =\left( \begin{matrix} 1 & -1 \\ -1 & 1 \end{matrix} \right)$, where the subindex $i$ refers to the subspace of velocities $\{\pm v_i\}$. The noise matrix $\Gamma_i$ is diagonalized with the matrix $O_i=\left(\begin{matrix} 1 & 1 \\ 1& -1 \end{matrix}\right)$ revealing a zero mode corresponding to $\delta \rho^+_i=\delta\rho^R_{i}+\delta\rho^L_{i}$, with $\delta\rho^{R/L}_{i}$ denoting the density of particles (above the background state) moving with velocity $\pm v_i$, respectively. There is also a decaying mode, corresponding to $\delta\rho^{-}_{i}=\delta\rho^R_{i}-\delta\rho^L_{i}$. Note that contrary to the non-interacting case, these {\em are not} normal modes of the hydrodynamic equations, since they do not diagonalize the velocity matrix $A$.
The diffusion matrix is thus given as
\begin{align}
\mathcal{D}_{i,j}&=\sum_k A_{(+,i),(-,k)}\Gamma^{-1}_{(-,k),(-,k)}A_{(-,k),(+,j)},
\end{align}
where the different matrices are written in the basis of $\pm$ modes (i.e. the matrix $A$ results from a rotation by $O=O_1\oplus O_2$). The resulting diffusion matrix has off-diagonal components, where some of these elements may be negative \cite{suppmat}. However, the matrix has strictly positive eigenvalues given by 
\begin{equation}\label{eq_diff_constant}
\mathcal{D}_{i}=\frac{(v_i^{\rm eff})^2}{2\gamma}, \hspace{0.05in} v_i^{\rm eff}=\frac{v_i}{1-an}, \hspace{0.05in} i=1,2.
\end{equation} 
Thus, the diffusion constant of the long-lived modes of the model, which are different from the conserved modes $\rho^{+}_{i}$ since the diffusion matrix is not diagonal (in contrast to the free particle case discussed above), is solely determined by the effective velocity of the original modes (in the integrable limit) and by the backscattering rate. This formula is also consistent with previous findings in the Rule 54 cellular automaton \cite{Lopez-Piqueres_2022}, and is analogous to the free particle case discussed above when replacing the velocities by their renormalized counterparts. This result is fairly intuitive: backscattering acts simply on the effective quasiparticles of the interacting model, so the mean-free path is set by the {\em effective} velocity instead of the bare one; we will come back to this intuition below. 

\begin{figure}
\includegraphics[scale=0.5, width=\linewidth]{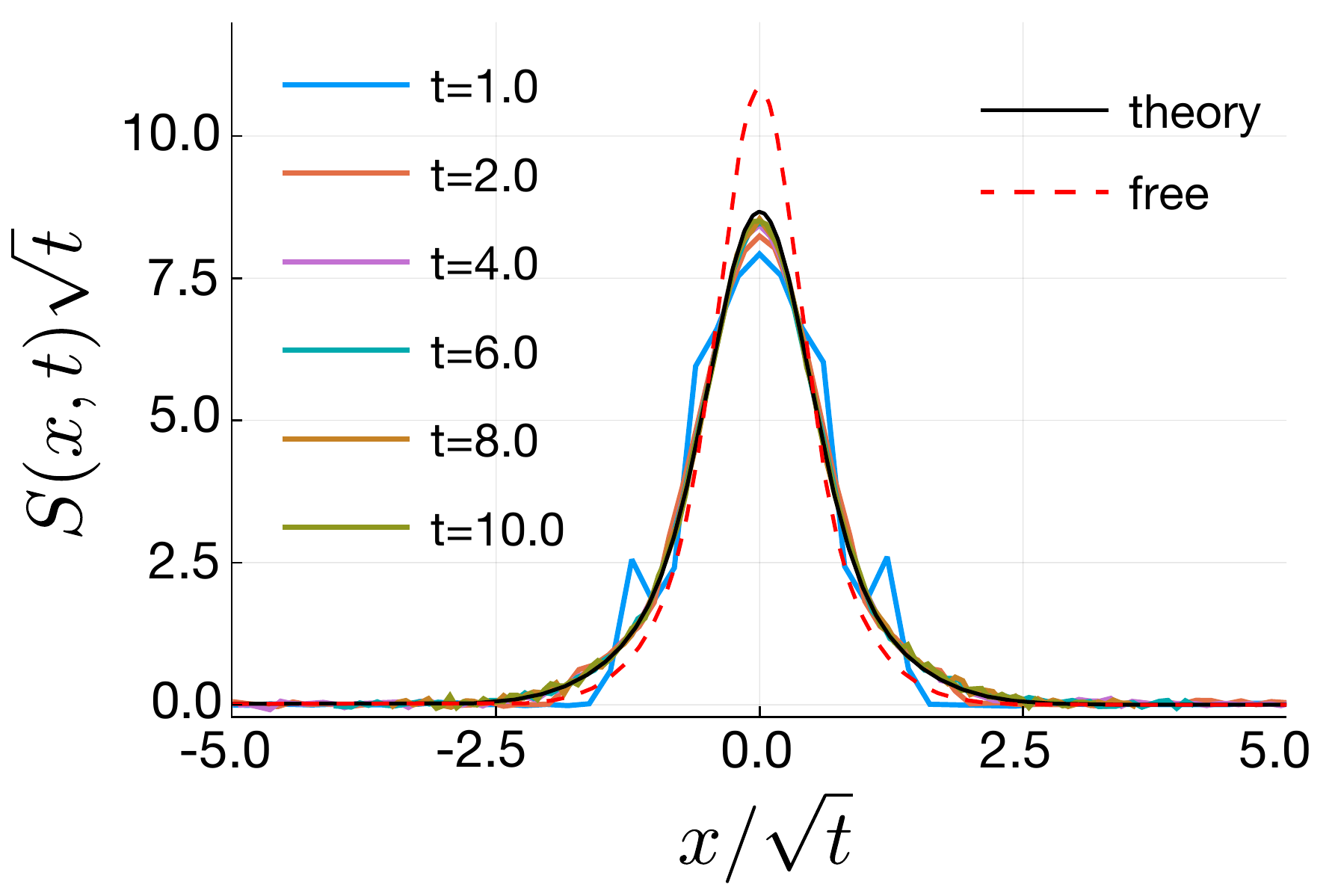}
\caption{\textbf{Structure factor(discrete case).} The background state is given by a uniform superposition of states with velocities $v_1=1$, $v_2=1/2$ (and $p_1=p_2=1/2$). The theory predictions follow Eq. (\ref{eq_struct_factor_uniform}) with the respective diffusion constants $\mathcal{D}_1$, $\mathcal{D}_2$. For comparison we also show the free theory results, corresponding to the $a\to 0$.} 
\label{fig_struct_factor_uniform}
\end{figure}

We focus now on the structure factor of the density of particles which is the observable of interest, giving us access to diffusion constant and a.c.~conductivities. This reads $S(x,t)=\langle \delta \rho(x,t) \delta \rho(0,0) \rangle_c$, with $\delta \rho = \delta \rho^{+}_{1}+\delta \rho^+_{2}$ and the label $c$ refers to the connected part of the correlator. With the aid of the eigenvector matrix that diagonalizes $\mathcal{D}$ given by $W$ with components $W_{i,i}=1-anp_i$, $W_{i,j\neq i}=-anp_i$, and the equilibrium charge fluctuation matrix $C=\langle \delta \rho \delta \rho \rangle = R^{-1}\rho^*R^{T}$ in the eigenmode basis, we can compute the structure factor matrix for the conserved modes $S_{i,j}(x,t) = \langle \delta \rho^+_i(x,t)\delta \rho^+_{j}(0,0)\rangle_c$. The exact expressions for these may be found in the Supp. Mat. \cite{suppmat}. The rod density structure factor is then given as $S(x,t)=\sum_{i,j}S_{i,j}(x,t)$, and we find the simple expression
\begin{equation} \label{eq_struct_factor_uniform}
S(x,t)=n(1-an)^{2}\langle g(x,2\mathcal{D}_it)\rangle,
\end{equation}
where $\langle \cdot \rangle = \sum_i p_i \cdot$ and $g(x,\sigma^2) =\frac{e^{-\frac{x^2}{2\sigma^2}}}{\sqrt{2\pi \sigma^2}}$. This expression is also consistent with the sum rule $\int dx S(x,t) = \sum_{i,j} C_{i,j}$. In Fig. \ref{fig_struct_factor_uniform} we present the results from simulating numerically the hard-rod gas where rods take in velocities $v_1=1$, $v_2=1/2$ with probabilities $p_1=p_2=1/2$.  The parameters used in the simulation are: backscattering rate $\gamma=0.005$, system size $2L=20$, number of hard-rods $N=400$, and hard-rod length $a=0.01$. We use periodic boundary conditions (pbc), and subtract off initial fluctuations due to finite size effects. For comparison we also present the results from the free theory predictions, corresponding to the limit $a \to 0$, showing that the dynamics is both chaotic and interacting. The small discrepancies from the theory predictions are the result of the dynamics not having fully thermalized on the timescales of the simulation.

\textbf{\textit{General case---}} When the spectrum of velocities is continuous, for instance, given by a Gaussian packet centered around $k=0$, the approach taken for a discrete spectrum is still helpful. Indeed it is straightforward to extend the previous analysis to the case of an arbitrary discrete spectrum of velocities by induction from the studied case of two particle species \cite{suppmat}. In particular, the diffusion constant of each of the hydrodynamic modes in the presence of backscattering will be given by Eq. (\ref{eq_diff_constant}). This result still carries over to the continuum. 

The tractability of this problem can be understood in terms of the simple action of the backscattering perturbation in terms of the normal modes of GHD, that is, the modes $\delta \tilde{\rho} = R \delta \rho$ that diagonalize the matrix $A$. Formally, the problem is dramatically simplified by the fact that $[R,\Gamma]=0$, where $R=1-\theta^*T$ is the matrix that diagonalizes $A$ (whose eigenvalues correspond to the effective velocities). The physical meaning of this constraint is that effectively, backscattering noise acts simply on the quasiparticles dressed by interactions.
In that  basis, the Boltzmann equation~(\ref{eq_BE}) now reads                    
\begin{equation} \label{eq_BE_interactions}
\partial_t \delta \tilde{\rho}_k+v_k^{\rm eff}\delta \tilde{\rho}_k=-\gamma(\delta \tilde{\rho}_k-\delta \tilde{\rho}_{-k}),
\end{equation}
where $\delta \tilde{\rho}_k \equiv (R\delta \rho)_k$ and $v_k^{\rm eff} = v_k^{\rm eff}[\rho^*]$. We note that this simplification occurs only if the backscattering rate is velocity independent, since $(T\gamma)_k \neq (\gamma T)_k$ in general. Further, the requirement that $(R\delta \rho)_{-k} = \delta \tilde{\rho}_{-k}$ follows from the equilibrium occupation number being an even function $\theta^*_{k}=\theta^*_{-k}$ (as it should in equilibrium), and from the symmetry of the scattering kernel $T_{k,-k'}= T_{-k,k'}$. The generalized Boltzmann equation (\ref{eq_BE_interactions}) is a direct generalization of eq.~(\ref{eq_BE}) in the presence of interactions, where the effective velocities are now dressed by the effects of interactions. The problem therefore reduces to the non-interacting one~\eqref{eqfreeBE}: backscattering leads to a $2\times2$ problem in the $(k,-k)$ basis of GHD normal modes. The residual hydrodynamic modes $\delta \tilde{\rho}^+_{k} \equiv \delta \tilde{\rho}_k + \delta\tilde{\rho}_{-k}$ satisfy
\begin{equation} \label{eq_PDE}
(\partial_t^2+2\gamma \partial_t - (v_k^{\rm eff})^2\partial_x^2)\delta\tilde{\rho}^+_{k}=0,
\end{equation}
which exhibits a crossover from ballistic transport at short times ($\gamma t  \ll 1$), to diffusive transport with diffusion constant $\mathcal{D}_k=(v_k^{\rm eff})^2/2\gamma$ at long times  ($\gamma t \gg 1$). Diffusion is induced by the decay of the nonconserved charges (`$-$' modes) with decay rate $2\gamma$.

\begin{figure}
\includegraphics[scale=0.5, width=\linewidth]{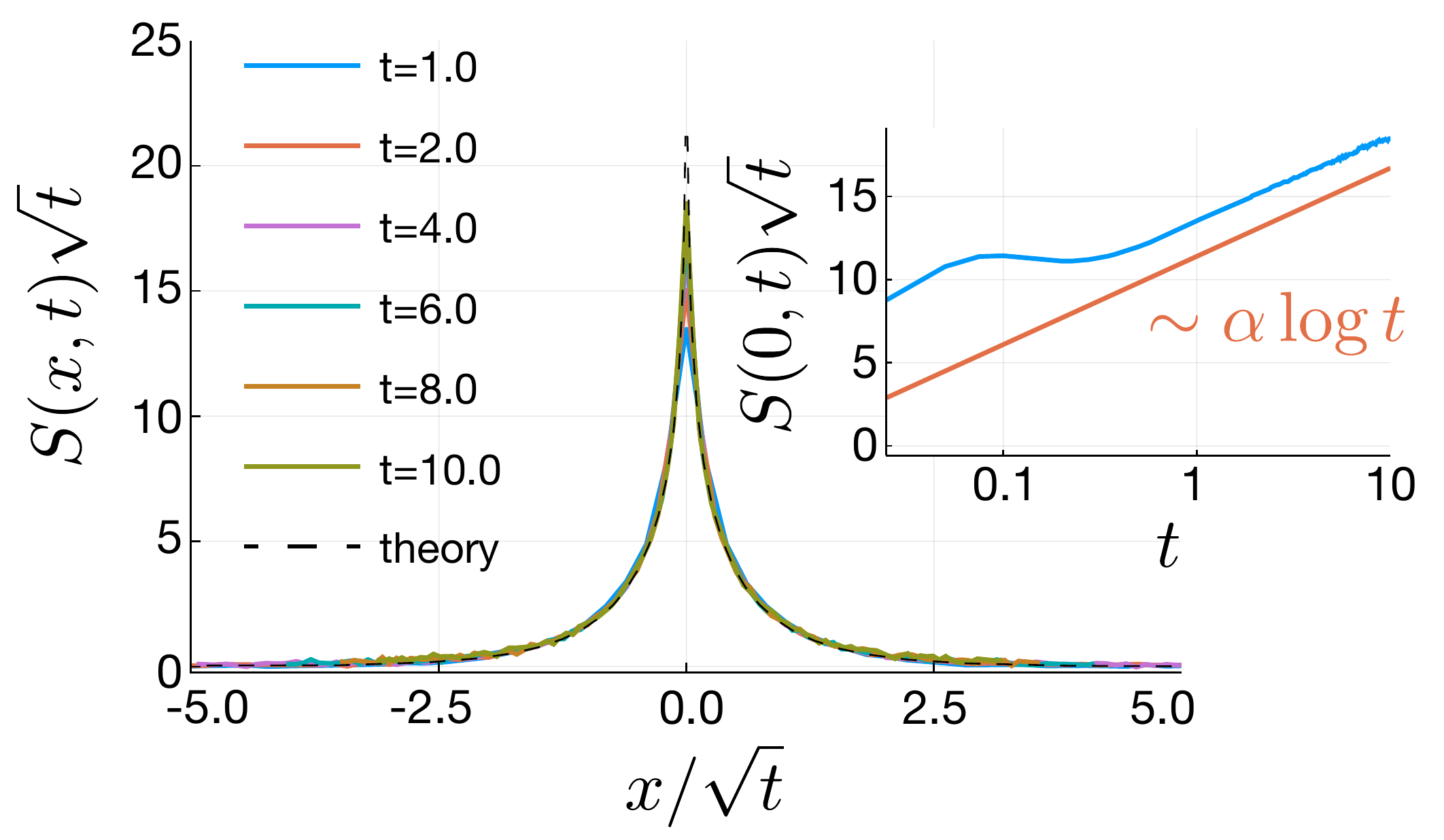}
\caption{\textbf{Anomalous structure factor.} The background state is given by a a Gaussian (thermal) velocity distribution centered at $0$ and variance $\sigma^2$. The theory predictions follow Eq. (\ref{eq_Bessel}). Inset: scaling of structure factor at $x=0$ (return probability) along with theory predictions (ignoring an offset for visual purposes), where $\alpha=n(1-an)^3/2\pi \sigma$, follows also from Eq. (\ref{eq_Bessel}). } 
\label{fig_struct_factor_gaussian}
\end{figure}

\textbf{\textit{Anomalous structure factor---}} Focusing on in the long time limit of the conserved modes, the resulting structure factor follows from that in Eq.~(\ref{eq_struct_factor_uniform}), with $n p_i \to \rho^*_k$ the hard-rod phase space density at equilibrium. Taking $\rho^*_k=n p(k)$ with $p(k)$ a Gaussian (thermal) velocity distribution centered at $0$ and with variance $\sigma^2$, the rod density structure factor reads
\begin{equation} \label{eq_Bessel}
S(x,t)=\frac{n(1-an)^3}{\pi \sigma}\sqrt{\frac{\gamma}{t}}K_0\left(\frac{1-an}{\sigma}\sqrt{\frac{\gamma}{t}}|x|\right),
\end{equation}
with $K_0(x) = \int_0^\infty {\rm e}^{- |x|\cosh t} dt$ the modified Bessel of second kind. In Fig.~\ref{fig_struct_factor_gaussian} we compare the theory predictions with the numerical results showing excellent agreement. We trace back this singular behavior to the presence of infinitely many conserved charges, each with a different diffusion constant, conspiring to produce a profile that is evidently nongaussian. In particular, the structure factor shows a singularity of logarithmic nature at the origin independently of the rods' length, following from $K_0(a x)\underset{x \to 0}{=}-\gamma_{\rm Euler}-\log(ax/2) + \mathcal{O}(x^2\log x)$, $a>0$, with $\gamma_{\rm Euler}$ Euler's constant. This implies that the {\em return probability} (structure factor near the origin) is {\em anomalous}, with a logarithmic correction to the expected diffusive behavior
\begin{equation} \label{eq_ReturnProba}
S(x\approx 0,t) \sim \frac{\log t}{\sqrt{t}},
\end{equation}
which we also observe in numerical simulations (Fig.~\ref{fig_struct_factor_gaussian}). 
The effective diffusion constant of hard-rods in this limit is found as $\mathcal{D}=\frac{1}{2t}\int dx x^2 S(x,t)$ which yields $\mathcal{D}=n\sigma^2/2\gamma$, independently of the rods' length. 

\textbf{\textit{Conclusion ---}} In this work we have explored the effects of backscattering noise in the hard-rod gas. We find that the density of rods spreads diffusively as a linear combination of Gaussians of different widths, corresponding to the different diffusion constants of the normal modes of the hydrodynamic theory. For a thermal velocity distribution, this leads to a singular structure factor with a logarithmic correction to the return probability. 
Our results generalize to other similar integrable models, such as the Lieb-Liniger model, so long as $[R,\Gamma]=0$, with the important caveat that the backscattering operator acts on the system's quasiparticles, not on the physical particles. Understanding better the relationship between these two backscattering sources and the relevance of backscattering in experimental setups of strongly interacting, confined Bose gases is left as future work. Another future extension of our work would be to study diffusive corrections to backscattering that arise from the integrable dynamics itself (\textit{i.e.} incorporating Navier-Stokes corrections~\cite{lebowitz1968time, spohn1982hydrodynamical, boldrighini1997one, doyon2017dynamics,PhysRevLett.121.160603,gopalakrishnan2018hydrodynamics,de2019diffusion} to the GHD equation~(\ref{eq_BE})). In this case, the expectation is that such diffusive and higher order corrections will be subleading when compared to the contributions coming from backscattering, since the latter contribute $1/\gamma$ in the limit of small noise, $\gamma \to 0$. A more interesting setup would be to study backscattering in the presence of an harmonic trap, which has been shown to break integrability~\cite{PhysRevLett.120.164101,PhysRevLett.125.240604}. The harmonic trap introduces a new timescale after which the system thermalizes. This timescale is anomalously large and it would be interesting to see whether backscattering can speed this up by breaking all odd charges in the reachable timescales seen in this work. 

\textit{Acknowledgements.} We thank Sarang Gopalakrishnan for helpful discussions and collaborations on related topics. This work was supported by the US Department of Energy, Office of Science, Basic Energy Sciences, under Early Career Award No. DE-SC0019168, and the Alfred P. Sloan Foundation through a Sloan Research Fellowship (R.V.).

\bibliography{refs}

\bigskip

\onecolumngrid
\newpage



\section*{Supplementary material}

# Supplemental Material for "Integrability breaking from backscattering"

Javier Lopez-Piqueres[1] and Romain Vasseur[1]

[1]*Department of Physics, University of Massachusetts, Amherst, Massachusetts 01003, USA*

## I. DIFFUSIVE HYDRODYNAMICS FROM BACKSCATTERING

Here we rederive the formula for the diffusion constant of conserved modes within GHD used in the main text, Eq. (4), and which follows from Ref.[1] (see also[2]). Our starting point is the linearized Boltzmann equation Eq. (3),

$$\partial_t \delta\rho + A \partial_x \delta\rho = -\Gamma \delta\rho, \tag{1}$$

where the operators $A$ and $\Gamma$ act on velocity space as convolution, e.g. $A\delta\rho_k(x,t) = \int dk' A_{kk'} \delta\rho_{k'}(x,t)$. The densities of conserved charges in the hard-rod gas obey

$$q_n(x,t) = \int dk\, k^n \rho_k(x,t). \tag{2}$$

Similar relations would still hold for other integrable models; in particular, an identical relation holds in the Lieb-Liniger model – see e.g. Ref.[3]. We can recast the Boltzmann equation 1 in terms of these charges as follows,

$$\partial_t \delta q_n + \sum_m A_{nm} \partial_x \delta q_m = -\sum_m \Gamma_{nm} \delta q_m, \tag{3}$$

where, analogously to the phase-space density, we assumed the above dynamics occurs above a state of fixed, homogeneous charge $q_n^* = \int dk\, k^n \rho_k^*$. Now the hydrodynamic matrices are expressed in the charge basis and have components e.g. $A_{nm} = \int dk\, dk'\, k^n A_{k,k'} k'^m$. We now use the method of hydrodynamic projections to project Eq. (3) onto decaying charges, via the operator $P$, and conserved charges, via its complement, which we denote as $\bar{P} = 1 - P$. This results in the following two equations

$$\partial_t \bar{P} \delta\vec{q} + \bar{P} A \partial_x \delta\vec{q} = 0, \tag{4a}$$
$$\partial_t P \delta\vec{q} + P A \partial_x \delta\vec{q} = -P \Gamma P \delta\vec{q}. \tag{4b}$$

where the charge vector $\vec{q}$ has as components all the charges of the original integrable model. Note that we have implicitly assumed that dissipation will only act within the subspace of decaying charge (while in general, and in particular, for backscattering, decaying and conserved modes will be mixed by the integrable dynamics). Applying the inverse of the dissipation operator on (4b) and solving recursively yields

$$\begin{aligned} P\delta\vec{q} &= -(P\Gamma P)^{-1}(\partial_t P \delta\vec{q} + PA\partial_x \delta\vec{q}) \\ &= -\partial_x (P\Gamma P)^{-1} PA\bar{P}\delta\vec{q} + \mathcal{O}(\{\partial_x^2, \partial_t\partial_x, \partial_t^2\}). \end{aligned} \tag{5}$$

Ignoring higher order gradient terms and plugging this into (4a) leads to

$$\partial_t \bar{P}\delta\vec{q} + \bar{P}A\partial_x \bar{P}\delta\vec{q} = \mathcal{D}\partial_x^2 \bar{P}\delta\vec{q}, \tag{6}$$

where

$$\mathcal{D} = \bar{P}A(P\Gamma P)^{-1}A\bar{P}. \tag{7}$$

We will show below that for the case of backscattering noise, there are no ballistic contributions as the matrix $A$ will only couple decaying charges with conserved ones.

## II. BACKSCATTERING IN THE HARD-ROD GAS

While the hydrodynamics of the hard-rod gas was originally solved assuming a (continuous) distribution of rods' velocities, we shall assume that a well-defined and analogous description exists in the presence of a discrete set of different velocities. This will facilitate computing the diffusion constant of the different particle species exactly, and in the limit of infinite number of such particle species, allow us to compute the structure factor of density in the continuum limit.

### A. One particle species (two velocities)

Earlier works[4–6] had shown that the hydrodynamics of the hard-rod gas at the Euler scale obeys GHD equation

$$\partial_t \rho + \partial_x (v^{\text{eff}} \rho) = 0, \tag{8}$$

where

$$v_k^{\text{eff}}[\rho] = k + \frac{a \int_{k'} (k - k') \rho_{k'}(x,t)}{1 - a \int_{k'} \rho_{k'}(x,t)}, \tag{9}$$

with $a$ the hard-rod's length. Within linear response we may take $\rho = \rho^* + \delta\rho$, with $\rho^*$ some stationary state that is spatially homogeneous. For what follows we take this state to be given by $\rho^* = n\frac{1}{2}(\delta(v - v_0) + \delta(v + v_0))$, with $n = N/(2L)$ the density of hard-rods in the system. In this stationary state the effective velocity is given by

$$v_{v_0}^{\text{eff}}[\rho^*] = \frac{v_0}{1 - an} \equiv v^{\text{eff},*}. \tag{10}$$

Note that this effective velocity is precisely that of the Rule 54 cellular automaton when setting $v_0 = \pm 1$, $a = -1$[7]. Carrying this analogy further, we posit the GHD equation to be of the form:

$$\partial_t \delta \vec{\rho} + A \partial_x \delta \vec{\rho} = 0, \tag{11}$$

with $\vec{\rho} = (\rho_+, \ \rho_-)$, with $\rho_\pm = \rho_R \pm \rho_L$, where $\rho_{R/L}$ is the density of particles moving at speed $v_0 = \pm 1$. In the linear response setup we're interested in we have $\rho_+ = n + \delta\rho$, $\rho_- = \delta\rho_-$, and where the operator $A$ is given as[6]

$$A = (1 - \theta^* T)^{-1} v^{\text{eff},*} (1 - \theta^* T), \tag{12}$$

where $\theta$ is the Fermi factor and we have implicitly assumed all quantities here are evaluated at the stationary state. The Fermi factor in vector form is given as

$$\vec{\theta}^* = (1 - an)^{-1} \vec{\rho}^*, \quad \vec{\rho}^* = (n, \ 0). \tag{13}$$

The counterpart in matrix form corresponds to the diagonal matrix constructed out of this vector. Likewise, we may view the dressed velocity as being a diagonal matrix with entries given by (10). The kernel acts as follows on a test function

$$T\psi(v) = -a \int dw \, \psi(w). \tag{14}$$

In the $R/L$ basis the kernel should read[8]

$$T = -a \begin{pmatrix} 1 & 1 \\ 1 & 1 \end{pmatrix}. \tag{15}$$

In the presence of backscattering noise, the GHD equation takes the form

$$\partial_t \delta \vec{\rho} + A \partial_x \delta \vec{\rho} = -\Gamma \delta \vec{\rho}, \tag{16}$$

with, the decay matrix

$$\Gamma = \gamma \begin{pmatrix} 1 & -1 \\ -1 & 1 \end{pmatrix}, \tag{17}$$

with $p$ the proba for a right/left mover to veer direction. Note that $\Gamma$ has one zero mode and one decaying mode, which is found upon rotating $\Gamma$ to the $\pm$ basis as

$$\Gamma \to O\Gamma O^{-1}, \tag{18}$$

with

$$O = \begin{pmatrix} 1 & 1 \\ 1 & -1 \end{pmatrix}. \tag{19}$$

The decaying mode corresponds to the $\rho_-$ mode and has eigenvalue $\Gamma_{-,-} = 2\gamma$. To get the diffusion constant of positive movers we also rotate the $A$ matrices onto the $+/-$ basis to find

$$A_{+,-} = 1, \quad A_{-,+} = \frac{1}{(1-an)^2}. \tag{20}$$

The diffusion constant is given as

$$\mathcal{D}_{+,+} = A_{+,-}\Gamma^{-1}_{-,-}A_{-,+} = \frac{1}{2\gamma(1-an)^2} \equiv \frac{(v^{\text{eff}})^2}{2\gamma}. \tag{21}$$

Note that these are results are identical to those obtained in Rule 54 upon setting $a = -1^9$.

### B. Two particle species

The previous analysis can be generalized straightforwardly to the case of particles having four possible velocities (two of opposite sign) with backscattering. The corresponding hydrodynamic matrices now read

$$T = -a \begin{pmatrix} 1 & 1 & 1 & 1 \\ 1 & 1 & 1 & 1 \\ 1 & 1 & 1 & 1 \\ 1 & 1 & 1 & 1 \end{pmatrix}, \tag{22}$$

$$\Gamma = \gamma \begin{pmatrix} 1 & -1 & 0 & 0 \\ -1 & 1 & 0 & 0 \\ 0 & 0 & 1 & -1 \\ 0 & 0 & -1 & 1 \end{pmatrix}, \tag{23}$$

$$O = \begin{pmatrix} 1 & 1 & 0 & 0 \\ 1 & -1 & 0 & 0 \\ 0 & 0 & 1 & 1 \\ 0 & 0 & 1 & -1 \end{pmatrix}. \tag{24}$$

where the rotation matrix is defined so as to rotate to the $\pm$ basis within each speed sector. We stick to the convention that the top left block in the $O$ matrix defines a subspace of e.g. velocity $\pm v_1$ and the bottom right block defines a different subspace of velocity $\pm v_2$. The Fermi factor (in matrix form) reads in the $R/L$ basis

$$\theta = \frac{1}{1-an}\frac{n}{4}\mathbb{1}_{4\times 4}, \tag{25}$$

where we've chosen the background state to be that where all velocities are equally likely to simplify the analysis (we will extend to arbitrary distribution and number of particle species below). With this we can find the $A$ matrix in

the $\pm$ basis using (12). For reference we write it down

$$A = \frac{1}{1-an}\begin{pmatrix} 0 & (1-\frac{an}{2})v_1 & 0 & -\frac{an}{2}v_2 \\ \frac{1-\frac{an}{2}}{1-an}v_1 & 0 & \frac{an}{2(1-an)}v_1 & 0 \\ 0 & -\frac{an}{2}v_1 & 0 & (1-\frac{an}{2})v_2 \\ \frac{an}{2(1-an)}v_2 & 0 & \frac{1-\frac{an}{2}}{1-an}v_2 & 0 \end{pmatrix}\begin{matrix}(+,1)\\(-,1)\\(+,2)\\(-,2)\end{matrix} \quad (26)$$

with $(\pm, i)$ corresponding to the sector with $\pm v_i$ velocity. The diffusion matrix written in this basis reads

$$\mathcal{D}_{(+,i),(+,j)} = \sum_k A_{(+,i),(-,k)} \Gamma^{-1}_{(-,k),(-,k)} A_{(-,k),(+,j)}, \quad (27)$$

with $\Gamma_{(-,i),(-,i)} = 2\gamma$. The diffusion matrix has off-diagonal entries now owing to the interaction between the two different particle species. The matrix components read

$$\mathcal{D}_{(+,i),(+,j)} = \frac{1}{2\gamma}\frac{1}{4(1-an)^3} \times \begin{cases} (2-an)^2 v_1^2 - (anv_2)^2, & i=j=1, \\ (2-an)^2 v_2^2 - (anv_1)^2, & i=j=2, \\ an(2-an)(v_i^2 - v_j^2), & i \neq j. \end{cases} \quad (28)$$

The diagonal components of this matrix may be negative for some choice of parameters. This is fine so long as $\mathcal{D} \geq 0$. This can be checked by diagonalizing $\mathcal{D}$ and checking that its eigenvalues are nonnegative. We've confirmed this and found the eigenvalues to be $\{(v_1^{\text{eff}})^2, (v_2^{\text{eff}})^2\} \times 1/2\gamma$. The corresponding eigenmodes read $\{(-\phi, 1), (1, -\phi)\}$ with $\phi \equiv -1 + \frac{2}{an}$. This crucially implies $\rho_+ \equiv \rho_{(+,1)} + \rho_{(+,2)}$ is not a diffusive mode. Instead, these correspond to

$$\delta\vec{\tilde{\rho}} = W^{-1}\delta\vec{\rho}, \quad (29)$$

with $\delta\vec{\rho} = \begin{pmatrix} \delta\rho_{(+,1)} & \delta\rho_{(+,2)} \end{pmatrix}$, and

$$W = \begin{pmatrix} -\phi & 1 \\ 1 & -\phi \end{pmatrix}. \quad (30)$$

the matrix of eigenvectors of $\mathcal{D}$. We are interested in the quantity $S(x,t) = \langle \delta\rho(x,t)\delta\rho(0,0)\rangle^c$, with $\delta\rho = \delta\rho_{(+,1)} + \delta\rho_{(+,2)}$ corresponding to the density of particles. First we compute the $C$ matrix which reads[6]

$$C = (1-\theta^*T)^{-1}\rho^*(1-T\theta^*)^{-1}. \quad (31)$$

At variance with the $A$ matrix, the $C$ matrix, written in the $+, -$ basis is *not simply* given by a rotation of $C$ by $O$, and instead this picks up a factor of 2 (as can be checked by explicitly computing $C$ and $A$). Explicitly

$$C = \begin{pmatrix} \frac{1}{4}n(2-2an+(an)^2) & 0 & \frac{1}{4}an^2(-2+an) & 0 \\ 0 & \frac{n}{2} & 0 & 0 \\ \frac{1}{4}an^2(-2+an) & 0 & \frac{1}{4}n(2-2an+(an)^2) & 0 \\ 0 & 0 & 0 & \frac{n}{2} \end{pmatrix}\begin{matrix}(+,1)\\(-,1)\\(+,2)\\(-,2)\end{matrix} \quad (32)$$

Crucially, the $(+,1)$ and $(+,2)$ charges are correlated, while $(-,1)$ and $(-,2)$ are not. This means in particular we must take care of cross terms when expanding $\delta\rho = \delta\rho_{(+,1)} + \delta\rho_{(+,2)}$ in the structure factor. We compute next each term appearing in $S(x,t)$, $S(x,t) = S_{1,1}(x,t) + S_{1,2}(x,t) + S_{2,1}(x,t) + S_{2,2}(x,t)$, with $S_{i,j}(x,t) = \langle \delta\rho_{(+,i)}(x,t)\delta\rho_{(+,j)}(0,0)\rangle^c$. We shall be interested in the $C$ matrix projected onto the $\{(+,1),(+,2)\}$ subspace. Call it $C^{(+)}$. In the eigenmode basis this reads

$$\tilde{C}^{(+)} = (W^{-1})C^{(+)}(W^{-1})^T = \frac{n}{2}\frac{(an)^2}{4}\mathbb{1}_{2\times 2} = \frac{n}{4}\frac{1}{1+\phi}\mathbb{1}_{2\times 2}. \quad (33)$$

Note that the exact expression of $\tilde{C}$ depends on the choice of normalization factors of the eigenvectors forming matrix $W$ but this is the one that yields the simplest expression for $\tilde{C}$. Then,

$$\begin{aligned} S_{1,1}(x,t) &= \phi^2 \tilde{S}_{1,1}(x,t) - \phi \tilde{S}_{1,2}(x,t) - \phi \tilde{S}_{2,1}(x,t) + \tilde{S}_{2,2}(x,t) \\ &\asymp \phi^2 \tilde{C}_{1,1}^{(+)} g(x, 2\mathcal{D}_1 t) + \tilde{C}_{2,2}^{(+)} g(x, 2\mathcal{D}_2 t) \\ &= \frac{n}{4(1+\phi)} \left( \phi^2 g(x, 2\mathcal{D}_1 t) + g(x, 2\mathcal{D}_2 t) \right), \end{aligned} \tag{34}$$

where $\mathcal{D}_i = (v_i^{\text{eff}})^2/2\gamma$, and $\asymp$ denotes the limit $t \gg 1/\gamma$, and $g(x,\sigma^2) = \frac{e^{-\frac{x^2}{2\sigma^2}}}{\sqrt{2\pi\sigma^2}}$. As a sanity check we confirm $\int dx S_{1,1}(x,t) = C_{1,1}$. We can compute the rest of the terms contributing to $S(x,t)$ and we find

$$\begin{aligned} S_{1,2}(x,t) &\asymp -\frac{n\phi}{4(1+\phi)} (g(x, 2\mathcal{D}_1 t) + g(x, 2\mathcal{D}_2 t)), \\ S_{2,1}(x,t) &\asymp -\frac{n\phi}{4(1+\phi)} (g(x, 2\mathcal{D}_1 t) + g(x, 2\mathcal{D}_2 t)), \\ S_{2,2}(x,t) &\asymp \frac{n}{4(1+\phi)} (g(x, 2\mathcal{D}_1 t) + \phi^2 g(x, 2\mathcal{D}_2 t)). \end{aligned} \tag{35}$$

These results also fulfill the sum rule $\int dx S_{i,j}(x,t) = C_{i,j}$. Equipped with these results we get the structure factor of the density

$$\begin{aligned} S(x,t) &\asymp \frac{n}{2} \left( \frac{1-\phi^2}{1+\phi^2} \right)^2 (g(x, 2\mathcal{D}_1 t) + g(x, 2\mathcal{D}_2 t)) \\ &= \frac{n}{2} (1-an)^2 (g(x, 2\mathcal{D}_1 t) + g(x, 2\mathcal{D}_2 t)). \end{aligned} \tag{36}$$

So, indeed, we find that the structure factor of the density is given by an equal superposition of the two gaussians.

### C. $m$-particle species (arbitrary distribution)

For arbitrary number of particles species and arbitrary even velocity distributions, so that $\{v_i \to p_i/2, -v_i \to p_i/2\}$, with $p_i \in [0,1]$, the corresponding $A$ matrix components read (found by inspection from analyzing smaller instances)

$$A_{(+,i),(-,j)} = \begin{cases} \frac{-1+anp_i}{-1+an} v_i, & i = j \\ \frac{an}{-1+an} p_i v_j, & i \neq j. \end{cases} \tag{37}$$

$$A_{(-,i),(+,j)} = \begin{cases} \frac{1-an(1-p_i)}{(-1+an)^2} v_i, & i = j \\ \frac{an}{(-1+an)^2} p_i v_i, & i \neq j. \end{cases} \tag{38}$$

Which results in the diffusion matrix

$$\mathcal{D}_{(+,i),(+,i)} = \frac{1}{2\gamma(1-an)^3} \left( (1 - an + (an)^2 p_i) v_i^2 - (an)^2 p_i \langle v^2 \rangle \right), \tag{39}$$

$$\mathcal{D}_{(+,i),(+,j \neq i)} = \frac{anp_i}{2\gamma(1-an)^3} \left( v_i^2 + (1+an) v_j^2 - an \langle v^2 \rangle \right), \tag{40}$$

with $\langle v^2 \rangle = \sum_k p_k v_k^2$. Note that in the absence of interactions, the diffusion matrix is purely diagonal. The next step is to determine the eigenmode matrix that diagonalizes $\mathcal{D}$. This has the rather simple form

$$W_{i,j} = \begin{cases} 1 - anp_i, & i = j \\ -anp_i, & i \neq j. \end{cases} \tag{41}$$

The charge-charge matrix for arbitrary distribution has components

$$C_{(+,i),(+,j)} = \begin{cases} np_i(1 - anp_i + (an)^2 p_i), & i = j, \\ -(2 - an)an^2 p_i p_j, & i \neq j, \end{cases}$$
$$C_{(-,i),(-,j)} = np_i \delta_{i,j},$$
$$C_{(+,i),(-,j)} = 0 = C_{(-,i),(+,j)}.$$
(42)

Note that the sum rule in this case takes a very simple form

$$\sum_{i,j} C_{(+,i),(+,j)} = n(1 - an)^2. \tag{43}$$

Written in the basis of eigenmodes in the + subspace the charge-charge matrix reads

$$\tilde{C}^{(+)} = n(an)^2 \text{diag}(p_1, p_2, \cdots, p_m). \tag{44}$$

The structure factor matrix has components

$$S_{i,j}(x,t) = n(an)^2 \times \begin{cases} \left(p_i - \frac{1}{an}\right)^2 p_i g(x, \mathcal{D}_i t) + \sum_{k \neq i} p_i^2 p_k g(x, 2\mathcal{D}_k t), & i = j \\ p_i p_j \left(\left(p_i - \frac{1}{an}\right) g(x, 2\mathcal{D}_i t) + \left(p_j - \frac{1}{an}\right) g(x, 2\mathcal{D}_j t) + \sum_{k \neq i,j} p_k g(x, 2\mathcal{D}_k t)\right), & i \neq j. \end{cases} \tag{45}$$

After some lengthy algebra we find the structure factor of density $S(x,t) = \sum_{i,j} S_{i,j}(x,t)$

$$S(x,t) = n(1 - an)^2 \langle g(x, 2\mathcal{D}t) \rangle, \tag{46}$$

where

$$\langle g(x, 2\mathcal{D}t) \rangle \equiv \sum_i p_i g(x, 2\mathcal{D}_i t). \tag{47}$$

This is the expression quoted in the main text. This expression fulfills the sum rule, $\sum_x S(x,t) = \sum_{i,j} C^{(+)}_{i,j} = n(1-an)^2$. In the continuum limit we simply replace $p_i \to p(v)$ and the sum by an integral. I.e. the structure factor is given as

$$\boxed{S(x,t) = (1-an)^2 \int dv \rho^*(v) g(x, 2\mathcal{D}(v)t), \quad \mathcal{D}(v) \equiv \frac{(v^{\text{eff}})^2}{2\gamma},} \tag{48}$$

with $\rho^*(v) = p(v)n$ the background particle density.

---



\end{document}